\documentclass[journal,comsoc]{IEEEtran}
\usepackage[T1]{fontenc}

\ifCLASSINFOpdf
\else
\fi
\usepackage{amsmath}
\usepackage[cmintegrals]{newtxmath}
\usepackage{algorithm}
\usepackage{algpseudocode}
\usepackage{enumerate}
\usepackage{subfig}
\usepackage{graphicx}
\usepackage{cite}
\usepackage{color}

\newtheorem{definition}{Definition}
\newtheorem{theorem}{Theorem}

\newtheorem{remark}{Remark}

\begin{document}
%
\title{Learning for Matching Game in Cooperative D2D Communication with Incomplete Information }

\author{Yiling~Yuan, Tao~Yang~\IEEEmembership{Member,~IEEE,}
        Hui~Feng,~\IEEEmembership{Member,~IEEE,}
        and~Bo~Hu,~\IEEEmembership{Member,~IEEE}
 \thanks{Copyright (c) 2015 IEEE. Personal use of this material is permitted. However, permission to use this material for any other purposes must be obtained from the IEEE by sending a request to pubs-permissions@ieee.org.}
\thanks{ This work was supported in part by the NSF of China (Grant No. 71731004, No. 61501124), and in part by the National Key Research and Development Program of China (No.213). \emph{(Corresponding author: Tao Yang).}}
\thanks{Y. Yuan, T. Yang and H. Feng  are with the Research Center of Smart Networks and Systems, the Department of Electronic Engineering, Fudan University, Shanghai, China  (e-mail: yilingyuan13@fudan.edu.cn; taoyang@fudan.edu.cn; hfeng@fudan.edu.cn). }
\thanks{B. Hu is with the Research Center of Smart Networks and Systems, Department of Electronic Engineering, Fudan University, Shanghai, China, and also with the Key Laboratory of Electromagnetic WaveInformation (MoE), Fudan University, Shanghai 200433, China (e-mail: bohu@fudan.edu.cn).}
}

\maketitle
\begin{abstract} 
This paper considers a cooperative device-to-device (D2D) communication system, where the D2D transmitters (DTs) act as relays to assist cellular users (CUs) in exchange for the opportunities to use licensed spectrum. Based on the interaction of each D2D pair and each CU, we formulate the pairing problem between multiple CUs and multiple D2D pairs as a one-to-one matching game. Unlike most existing works, we consider a realistic scenario with incomplete channel information. Thus, each CU lacks enough information to establish its preference over D2D pairs. Therefore, traditional matching algorithms are not suitable for our scenario. To this end, we convert the matching game to an equivalent non-cooperative game, and then propose a novel learning algorithm, which converges to a stable matching.
\end{abstract}

\begin{IEEEkeywords}
Cooperative D2D communication, matching game, incomplete information.
\end{IEEEkeywords}

%
\IEEEpeerreviewmaketitle

\section{Introduction}
\IEEEPARstart{R}{ecently}, D2D communication has been extensively studied to provide better user experience. To implement this technology, one of the key issues is how to share licensed spectrum efficiently without degrading CUs' performance greatly. We consider a cooperative D2D communication scheme, which exploits the advantages of cooperative relay and D2D communication \cite{Cao2015}. The basic idea is that DTs act as relays for CUs in exchange for the transmission opportunities on the CUs' channels. Thus, a win-win situation is achieved, which motivates CUs to share their spectrum with D2D pairs even if they have no surplus resource.

\par Most existing works\cite{Cao2015,Wu2017,6952682, Seif2017IWCMC} assume complete information, such as channel state information (CSI). However, collecting global information incurs heavy overhead, and thus may be not practical in large-scale networks. Besides, some information may be difficult to acquire, such as the CSI between CUs and DTs. Moreover, the latency requirement of some applications is stringent, such as D2D-based \mbox{vehicle-to-vehicle} communications. These facts motivate us to study distributed resource allocation scheme with incomplete information, where agents make decisions independently based on local information.

\par Game theory provides a framework to study the interactions of autonomous agents. There have been many game theoretical solutions in D2D networks \cite{Song2014}. In our context, CUs have preferences over D2D pairs and vice versa. Matching theory offers a suitable tool to study the cooperation between competitive CUs and  competitive D2D pairs. There have been some matching-based resource allocation schemes for D2D communication\cite{Zhou2017,Bayat2016,Zhou2018}. In this paper, we formulate the problem of pairing CUs with D2D pairs as a one-to-one matching game to seek a stable matching.

\par In  the literature, authors of \cite{Ma2016TCOM} have considered the incomplete information scenario, but do not investigate the pairing problem. Besides, similar cooperative scheme has been studied in cognitive radio networks recently\cite{Jayaweera2011JSAC,Yan2013JSAC,Feng2015WCMC,Feng2014,Duan2014,Lopez2016}, where secondary users (SUs) relay primary users' (PUs) traffic for rewards of the transmission opportunities. Some works adopt auction\cite{Jayaweera2011JSAC}, dynamic Bayesian game\cite{Yan2013JSAC},  and Stackelberg game\cite{Feng2015WCMC} to tackle  the incomplete information. Moreover, the authors of \cite{Feng2014} consider the incomplete information in the matching game model. However, above works \cite{Jayaweera2011JSAC,Yan2013JSAC,Feng2015WCMC,Feng2014} assume PU has the knowledge of the relay rates, which depend on the SUs' local information. In practice, such information is usually not known globally. In this paper, we consider a stronger incomplete information scenario, where CUs have no knowledge of the relay rates provided by the D2D pairs.  The authors of \cite{Duan2014,Lopez2016} consider the similar information assumption, but only consider single PU case. Instead, we consider the case with multiple CUs and multiple D2D pairs.

\par This paper focuses on the uplink resource sharing with incomplete information, because mobile devices are more likely to need help due to limited power budget. We formulate the pairing problem  as a one-to-one matching game, based on the interaction between each CU and each D2D pair. Such interaction is described by Nash bargaining solution (NBS). Because the relay rates are unknown, CUs cannot establish preferences over D2D pairs. Thus, traditional matching algorithms, such as Gale-Shapley (GS) algorithm,  are not suitable for our scenario. To the best of our knowledge, it is the first attempt to address the matching game with \emph{unknown preference}. To overcome the difficulty, we convert the matching game to an equivalent non-cooperative game. At each period, each CU selects a D2D pair and a corresponding time allocation, and obtains a payoff as feedback. Based on the feedback, we propose a learning algorithm, which is proven to converge to a stable matching in probability. Moreover, the corresponding time allocation converges to the result of NBS with probability 1.

\section{System Model and Problem Formulation}
\subsection{System Model}
We consider uplink resource sharing of a single cell with a base station (BS) denoted by $b$ and $M$ CUs. The set of CUs is denoted by $\mathcal{M}$. Besides, there are $N$ D2D pairs, and the set of them is denoted by $\mathcal{N}$. Each D2D pair contains one DT and one D2D receiver (DR). In this paper, we assume $M\leq N$. However, the proposed algorithm can be applied to the case where $M> N$. CU $m$ has been assigned to one cellular channel, namely channel $m$. There is no dedicated channel for D2D pairs. Therefore, D2D pairs relay the uplink traffic in exchange for access to the cellular channels.

\begin{figure}
\centering
\includegraphics[width=3.3in]{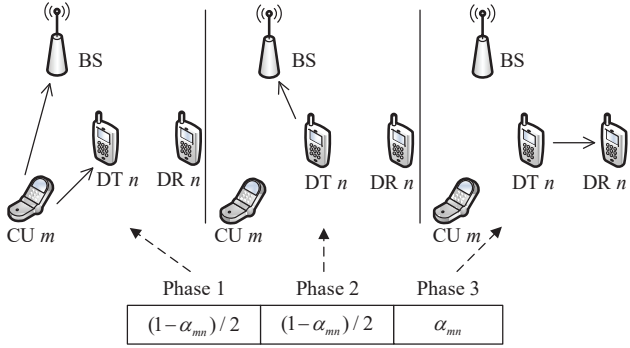}
\captionsetup{font={small}} 
\caption{Frame structure for cooperation.}
\label{frameStructure}
\end{figure}

\par
 We assume that each CU is assisted by at most one D2D pair, and each D2D pair can relay at most one CU due to limited battery capacity\cite{Wu2017}. Similar to \cite{Duan2014,Lopez2016, Feng2014}, we take the decode-and-forward protocol with parallel channel coding \cite{Laneman2003} as an example. When CU $m$ cooperates with D2D pair $n$, the normalized frame consists of  three phases, as shown in Fig.\ref{frameStructure}. The first two phases both last $\frac{1-\alpha_{mn}}{2}$ and are used for the relay transmission for the CU. Specifically, CU $m$ broadcasts its data with power $p_c$ to the BS and DT $n$ at first. Then, DT $n$ forwards received signal to the BS with power $p_d$. The third phase lasts $\alpha_{mn}$ and is used by DT $n$ to transmit its data with power $p_d$ to DR $n$. We refer to $\alpha_{mn}\in\mathcal{A}\triangleq[\alpha_{L},\alpha_{U}]$ as \emph{time allocation}.
\par
The expected rate of CU $m$ in direct link is
\begin{equation}\label{equ1}
  R^C_m =  \mathbb{E}\left[\ln\left(1+\frac{p_ch_{mb}}{n_0}\right)\right],
\end{equation}
where $h_{mb}$ is the channel gain from CU $m$ to the BS and $n_0$ denotes the noise power.

\par
For simplicity, we assume every DT can decode all the CUs' data in the first phase. Thus, cooperating with D2D pair $n$, the rate of CU $m$ in the first two phases is
\begin{equation}
r^C_{mn}=\frac{1}{2}\left[\ln\left(1+\frac{p_ch_{mb}}{n_0}\right)+\ln\left(1+\frac{p_dh_{nb}^m}{n_0}\right)\right],
\end{equation}
where $h_{nb}^m$ is the channel gain from DT $n$ to the BS on channel $m$. Let $R^C_{mn}=\mathbb{E}[r^C_{mn}]$, and thus with time allocation $\alpha_{mn}$, the expected rate of CU $m$ during the entire frame is $\tilde{R}^C_{mn}=(1-\alpha_{mn})R^C_{mn}$. Moreover, the expected rate of D2D pair $n$ during the entire frame is given by
\begin{equation}\label{equ3}
  \tilde{R}^D_{mn}(\alpha_{mn})=\alpha_{mn}\mathbb{E}\left[\ln\left(1+\frac{p_ch_{nn}^m}{n_0}\right)\right]\triangleq\alpha_{mn}{R}^D_{mn},
\end{equation}
where $h_{nn}^m$ is the channel gain of D2D pair $n$ on channel $m$. Assume that for each D2D link, the channel gains are i.i.d. across all the channels. Thus, we have ${R}^D_{mn}={R}^D_{m'n},\forall m,\forall m'\in\mathcal{M}$, and the value of ${R}^D_{mn}$ is denoted by $R^D_n$.

\par
\emph{Information Assumption:} CU $m$ only knows $R^C_{m}$ and has no knowledge of $R^C_{mn}$ and $R^D_{n}$, and D2D pair $n$ only knows $R^D_{n}$. After cooperating with D2D pair $n$ at period $t$, CU $m$ gets a sample $r^C_{mn}(t)$ following a fixed unknown distribution.

\subsection{Matching Based Framework}
\subsubsection{Bargaining Game for CU-D2D Pair $(m,n)$}
To incentivize CU and D2D pair to cooperate mutually, a bargaining game is used to characterize the interaction between them. If CU $m$ cooperates with D2D pair $n$, the CU's utility $U^C_{mn}$ and the D2D pair's utility $U^D_{mn}$ are defined as
\begin{align}
  U^C_{mn}(\alpha_{mn}) &=\tilde{R}^C_{mn}(\alpha_{mn})-R^C_m, \label{equ4} \\
  U^D_{mn}(\alpha_{mn}) &=\tilde{R}^D_{mn}(\alpha_{mn}). \label{equ5}
\end{align}
We use NBS as the bargaining outcome to determine the time allocation, and thus the cooperation satisfies some useful properties and is beneficial for both sides. Hence, based on the concept of NBS\cite{gametheory2011}, the time allocation is given by the following problem
\begin{subequations}\label{equ6}
\begin{align}
  \max_{\alpha_{mn}\in\mathcal{A}}   &\left(U^C_{mn}(\alpha_{mn})-U^C_{min}\right)\left(U^D_{mn}(\alpha_{mn})-U^D_{min}\right)           &\quad \label{equ6:a}\\
  \text{s.t.}\quad          &U^C_{mn}(\alpha_{mn})> U^C_{min}, U^D_{mn}(\alpha_{mn})> U^D_{min}, &\quad \label{equ6:b} 
\end{align}
\end{subequations}
where $U^C_{min}$ and $U^D_{min}$ are the CU's and the D2D pair's utilities respectively if they fail to reach an agreement. It is natural to set $U^C_{min}=U^D_{min}=0$. Thus, problem (\ref{equ6}) is coincident with proportional fairness scheme. Constraint (\ref{equ6:b}) guarantees that both sides have incentive to participate in the cooperation. Solving problem (\ref{equ6}), the optimal time allocation is given by
\begin{equation}\label{equ7}
  \alpha^*_{mn}(R^C_{mn})=\left[\frac{R^C_{mn}-R^C_m}{2R^C_{mn}}\right]_{\alpha_{L}}^{\alpha_{U}},
\end{equation}
where $[x]_a^b=\max(a,\min(x,b))$. Based on (\ref{equ7}), the D2D pair with higher relay rate can obtain larger transmission time. Moreover, it is easy to verify that $U^C_{mn}(\alpha^*_{mn}(R^C_{mn}))$ is an increasing function of $R^C_{mn}$, which reflects the fact that the CU prefers to cooperate with the D2D pair offering higher relay rate. We will use $\alpha_{mn}^*(R^C_{mn})$ and $\alpha_{mn}^*$ interchangeably afterwards. When the problem (\ref{equ6}) is infeasible, for convenience, we still let $\alpha_{mn}^*$ be the associated time allocation, and thus have $U^C_{mn}(\alpha_{mn}^*)\leq 0$ in this case. 

\subsubsection{Matching Game Model}
CU and D2D pair can only be paired when they agree to cooperate mutually. Therefore, it is reasonable to model the pairing problem between the set of CUs and the set of D2D pairs as a one-to-one matching game under two-sided preferences. CU $m$ prefers D2D pair $n$ to D2D pair $n'$ (i.e., $n\succ_m n'$), if $U^C_{mn}(\alpha_{mn}^*)>U^C_{mn'}(\alpha_{mn'}^*)$. Similarly, D2D pair $n$ prefers CU $m$ to CU $m'$ (i.e., $m\succ_n m'$), if $U^D_{mn}(\alpha_{mn}^*)>U^D_{m'n}(\alpha_{m'n}^*)$, which is equivalent to $\alpha_{mn}^*>\alpha_{m'n}^*$. Besides, if $U^C_{mn}(\alpha_{mn}^*)> 0$, D2D pair $n$ is \emph{acceptable} to CU $m$, which is denoted by $n\succ_m\emptyset$.
\par
Mathematically, a \emph{matching} is a function $\mu:\mathcal{M}\cup\mathcal{N}\rightarrow\mathcal{M}\cup\mathcal{N}\cup\{\emptyset\}$, such that $\mu(m)=n$ if and only if $\mu(n)=m$, and $\mu(m)\in\mathcal{N}\cup \{\emptyset\}$, $\mu(n)\in\mathcal{M}\cup\{\emptyset\}$, for $\forall m\in\mathcal{M},\forall n\in\mathcal{N}$. Note that $\mu(x)=\emptyset$ implies that user $x$ is unmatched. We aim to seek a \emph{stable matching} (SM), which is the major solution concept in matching game and defined as follows \cite{Roth1990Two}.

\begin{definition}\label{stableMatching}
Let $\mu$ be a matching. A CU-D2D pair $(m,n)$ is a \emph{blocking pair}  if $\mu(m)\neq n$, $m\succ_n\mu(n)$ and $n\succ_m\mu(m)$. $\mu$ is \emph{individually rational} if $\mu(m)\succ_m \emptyset,\forall m\in\mathcal{M}$. Thus, $\mu$ is \emph{stable} if it is individually rational and there is no blocking pair.
\end{definition}
\par SM captures the preferences of both sides and CUs will only be matched with acceptable D2D pairs in SM. The existence of SM is guaranteed\cite{Roth1990Two}. The challenge is that each CU cannot establish its preference due to the unavailability of $R^C_{mn}$. Thus, the traditional GS algorithm \cite{Roth1990Two} cannot be used to seek SMs.

\section{Learning for Matching with Incomplete Information}
To overcome the difficulty, CU has to learn its preference from the interactions with D2D pairs. To this end, we convert the above matching game  to an \emph{equivalent} non-cooperative game, which enables us to exploit the rich learning techniques designed for non-cooperative game.

\subsection{Equivalent Non-cooperative Game Model}
 We convert the matching game to a non-cooperative game $\mathcal{G}=(\mathcal{M},\{\mathcal{B}_m\}_{m\in\mathcal{M}},\{Ch_n\}_{n\in\mathcal{N}},\{\pi_m\}_{m\in\mathcal{M}})$. Due to the priority of CUs on licensed spectrum, we let CUs be the players to propose to D2D pairs. The action of CU $m$ is to select a D2D pair $b_m\in\mathcal{N}$, which {means CU $m$} proposes to cooperate with D2D pair $b_m$ with time {allocation $\alpha^*_{mb_m}$}. Each CU can refuse to cooperate with any D2D pairs, which is denoted by action $b_0$. Hence, the action set of CU $m$ is $\mathcal{B}_m=\mathcal{N}\cup\{b_0\}$. Given an action profile $\mathbf{b}=(b_1,b_2,\cdots,b_M)$, each D2D pair selects the CU offering the maximal time allocation among the CUs proposing to it and rejects the others. If more than one CUs offers the maximal time allocation, the D2D pair will choose one of them based on a predefined rule. The CU chosen by D2D pair $n$ is denoted by $Ch_n(\mathbf{b})$\footnote{Mathematically, the choice function of D2D pair $n$ can be represented as $Ch_n(\mathbf{b})=\arg\max_{m\in\mathcal{M}_n(\mathbf{b})}\{\alpha^*_{mn}+w_m\}$, where  $\mathcal{M}_n(\mathbf{b})$ is the set of CU proposing to D2D pair $n$  and $w_m$ is the bias assigned to CU $m$. The bias is determined by the predefined rule, and satisfies that if $\alpha^*_{mn}>\alpha^*_{m'n}$, $\alpha^*_{mn}+w_{m}>\alpha^*_{m'n}+w_{m'}$ must hold. }, which can reflect the preference of D2D pair $n$. Thus, the utility of CU $m$ is:
\begin{equation}\label{equ8}
  \pi_m(b_m,\mathbf{b}_{-m})=
  \begin{cases}
    U^C_{mb_m}(\alpha^*_{mb_m})-\theta, & \mbox{if }  Ch_{b_m}(\mathbf{b})= m,b_m\neq b_0,\\
    -\theta, & \mbox{if } Ch_{b_m}(\mathbf{b})\neq m,b_m\neq b_0, \\
    0, & b_m=b_0,
  \end{cases}
\end{equation}
where $\mathbf{b}_{-m}$ is the action profile of all the CUs except \mbox{CU $m$}, and $\theta>0$ is an arbitrarily small number and denotes the negotiation cost. Assume $\theta$ is sufficiently small so that $U^C_{mn}(\alpha^*_{mn})-\theta>0$ if $U^C_{mn}(\alpha^*_{mn})>0$. In the first case, $\theta$ makes sure that CUs only select acceptable D2D pairs at equilibriums. The first two cases imply acceptance and rejection of the CU's proposal, respectively. The third case means that the CU refuses to cooperate with any D2D pairs.

\par
Given an action profile $\mathbf{b}$, its associated matching $\mu_{\mathbf{b}}$ is obtained as follows: for $\forall m\in\mathcal{M},\forall n\in\mathcal{N}$, $\mu_{\mathbf{b}}(m)=n$ and $\mu_{\mathbf{b}}(n)=m$ if and only if $Ch_n(\mathbf{b})=m$. Hence, the relationship between the pure Nash equilibrium (PNE) of $\mathcal{G}$ and the SM can be stated as follows, which implies that an SM can be found via finding a PNE of $\mathcal{G}$.
\begin{theorem}
  If action profile $\mathbf{b}$ is a PNE, $\mu_\mathbf{b}$ is an SM. Conversely, if $\mu$ is an SM, there is a PNE $\mathbf{b}$ such that $\mu_{\mathbf{b}}=\mu$.
\end{theorem}

\begin{IEEEproof}
On the one hand, let $\mathbf{b}$ be a PNE.  We will prove the stability of $\mu_\mathbf{b}$ by contradiction. The individual rationality is easy to verify. Suppose  there is a blocking pair $(m,n)$ in $\mu_\mathbf{b}$. Thus, CU $m$ can take action $b'_m=n$ to improve its utility, which violates our assumption. Therefore, $\mu_{\mathbf{b}}$ is stable.

\par
On the other hand, let $\mu$ be an SM. We construct an action profile $\mathbf{b}$ as follows: for CU $m$, if $\mu(m)=n$, it takes action $b_m=n$ and action $b_0$ otherwise. We will prove that  $\mathbf{b}$ is  a PNE by contradiction. Suppose $\mathbf{b}$ is not a PNE, so there exists a CU $m$ deviating to take action $b'_m\neq b_m$. If $b'_m=b_0$, $\mu$ is not individually rational. Besides, If $b'_m\in\mathcal{N}$, there is a blocking pair in $\mu$. Thus, $\mu$ is not stable, which violates our assumption. Therefore, $\mathbf{b}$ is a PNE.
\end{IEEEproof}

\par
To develop the learning algorithm, we show that $\mathcal{G}$ is a \emph{weakly acyclic under better-replies game} (WABRG), which enables us to adopt \emph{better-reply with inertia} (BRI) learning algorithm \cite{chapman2013convergent}  to find the PNE of $\mathcal{G}$. WABRG means that from any action profile, there is a \emph{better-reply path} that terminates in a PNE in a finite number of steps. A \emph{better-reply path} is a sequence of action profiles $(\mathbf{b}^1,\mathbf{b}^2,\cdots,\mathbf{b}^t,\cdots)$, where for each $t$, there is a CU $m$ such that $b_m^t\neq b_m^{t-1}$, $\mathbf{b}_{-m}^t=\mathbf{b}_{-m}^{t-1}$ and $\pi_m(\mathbf{b}^t)>\pi_m(\mathbf{b}^{t-1})$. In other words, in successive action profiles, only one CU changes its action to improve its utility.
\begin{theorem}
The proposed game $\mathcal{G}$ is a WABRG.
\end{theorem}

\begin{IEEEproof}
Suppose $\mathbf{b}^0$ is not a PNE. We will construct a better-reply path that ends at a PNE to prove the theorem.
\par If there are any rejected CUs, we let them take action $b_0$ successively to obtain $\mathbf{b}^1,\mathbf{b}^2,\cdots,\mathbf{b}^{t_1}$, such that the CUs unmatched in $\mu_{\mathbf{b}^{t_1}}$ take action $b_0$. Furthermore, according to Theorem 2.33 in \cite{Roth1990Two}, there exists a finite sequence of matchings $\mu_1,\mu_2,\cdots,\mu_k,\cdots,\mu_K$, where $\mu_1=\mu_{\mathbf{b}^{t_1}}$, $\mu_K$ is stable, and there is a blocking pair $(m_k,n_k)$ for $\mu_k$ such that $\mu_{k+1}$ is obtained from $\mu_k$ by satisfying the blocking pair $(m_k,n_k)$. Thus, we let CU $m_1$ select D2D pair $n_1$ to obtain $\mathbf{b}^{t_1+1}$. Similarly, we let rejected CUs take action $b_0$ successively to obtain $\mathbf{b}^{t_1+2},\mathbf{b}^{t_1+2},\cdots,\mathbf{b}^{t_2}$. Note that the above process will not change the associated matching, i.e., $\mu_{\mathbf{b}^{t_2}}=\mu_2$. Repeating the above process, we can obtain an action profile $\mathbf{b}^{t_K}$ such that $\mu_{\mathbf{b}^{t_K}}=\mu_K$ and the unmatched CUs in $\mu_K$ take action $b_0$. Note that $\mathbf{b}^{t_K}$ is exactly the constructed action profile in the proof of Theorem 1, so it must be a PNE. Besides, it is easy to find that the sequence $\mathbf{b}^0,\mathbf{b}^1,\cdots,\mathbf{b}^{t_K}$ is a better-reply path. Hence, we can verify Theorem 2.
\end{IEEEproof}

\subsection{Learning Algorithm}
 Because each CU's utility is related to $R^C_{mn}$, each CU has to learn its utility from the interactions with D2D pairs. Furthermore, the action of CU $m$ can be redefined as a proposal $\hat{b}_m=(b_m,\alpha^*_{mb_m})$, where the time allocation is unknown in the case of incomplete information. Therefore, CUs have to make proposals explicitly to help D2D pairs establish their preferences. Specifically, at each period $t$, based on history information, CU $m$ makes proposal $(b^t_m,\alpha^t_{m})$, where $\alpha^t_{m}$ is calculated using the estimation of $R^C_{mb^t_m}$. Based on the proposal profile $\hat{\mathbf{b}}^t=(\hat{b}^t_1,\hat{b}^t_2,\cdots,\hat{b}^t_M)$, D2D pair $n$ selects the CU offering the maximal time allocation, and the selected CU is denoted by $\hat{Ch}_n(\hat{\mathbf{b}}^t)$. After cooperation with D2D pair $n$, CU $m$ can update its estimation using observation $r^C_{mn}$. Besides, to facilitate the learning process, CU $m$ can also choose the time allocation $\alpha_e=\alpha_{U}+\theta'$ to make sure it has enough chances to cooperate with every D2D pair to obtain information, where $\theta'>0$ is an arbitrary small number. Hence, with D2D pair $n$ selected, CU $m$ can choose $\alpha_e$ for exploration.

 \par
Combining BRI and Q-learning, we propose a novel learning algorithm. The entire algorithm is depicted in Algorithm 1 for some CU $m\in\mathcal{M}$.  In step 3, CU $m$ randomly selects D2D pair for exploration with probability $\varepsilon(t)$, where step 3-a is used to announce time allocation $\alpha^t_{mn}$ to help other CUs estimate their utilities. In step 4, with probability $1-\varepsilon(t)$, CU $m$ adopts BRI to learn a PNE of $\mathcal{G}$ using estimated utility $\hat{\pi}_m$. In step 6, based on observations, CU $m$ updates its estimation of  $R^C_{mn}$ in Q-learning way. Then, CU $m$ uses this updated estimation to calculate the associated time allocation in step 7. In step 9, CU $m$ uses other CUs' announced time allocation and $\alpha^t_{mn}$ to estimate utility function.

  \begin{algorithm}[t!]
 \caption{Extended BRI with Q-learning (EBRI-Q)}
 {\fontsize{8pt}{0.9\baselineskip}\selectfont
 \begin{algorithmic}[1]
 \State  Initialize $\hat{R}^C_{mn}(1),\hat{\alpha}^1_{m'n},\forall n\!\in\!\mathcal{N},\forall m'\!\in\!\mathcal{M}$ and $\hat{\pi}_m(\mathbf{b}),\forall\mathbf{b}\!\in\!\Pi_{m'\!\in\!\mathcal{M}}\mathcal{B}_{m'}$.
 \State \textbf{for} $t=2,3\cdots,T$
 \State With probability $\varepsilon(t)$, uniformly select D2D pairs $b^t_m$:
 \begin{enumerate}[a)]
 \item With probability $\zeta$, choose the time allocation as $\alpha^t_m=\alpha^t_{mb_m^t}$.
 \item With probability $1-\zeta$, choose the time allocation as $\alpha^t_m=\alpha_e$.
 \end{enumerate}
 \State With probability $1-\varepsilon(t)$, choose D2D pairs $b^t_m$ by following \emph{better reply with inertia} and choose $\alpha^t_m=\alpha^{t-1}_{mb_m^{t}}$:
 \begin{enumerate}[a)]
 \item With probability $\xi$, select D2D pair $b^t_m=b^{t-1}_m$.
 \item With probability $1-\xi$, select D2D pair $b^t_m$ according to the distribution, which is over the D2D pair selections that are better replies to CU's full memory of length $L$ than $b^{t-1}_m$ with respect to $\hat{\pi}_m$.
 \end{enumerate}
 \State Observe joint proposal $\hat{\mathbf{b}}^t$ and choice of each D2D pair. Get achieved rate $r^C_{mn}(t)$ if cooperating with D2D pair $n$.
 \State Update the estimation of $R^C_{mn},\forall n\in\mathcal{N}$:
 \begin{equation}\label{a1}
 \hat{R}^C_{mn}(t)=\hat{R}^C_{mn}(t-1)+\lambda(t)\mathbf{I}(\hat{Ch_n}(\hat{\mathbf{b}}^t)=m)(r^C_{mn}(t)-\hat{R}^C_{mn}(t-1)),
 \end{equation}
 where  $\lambda(t)={1}/{(1+\sum_{\tau=1}^t\mathbf{I}(\hat{Ch_n}(\hat{\mathbf{b}}^\tau)=m))}$ and $\mathbf{I}(\cdot)$ is indicator function.
 \State Update its time allocation: $\alpha^t_{mn}=\alpha^*_{mn}(R^C_{mn}(t)),\forall n\in\mathcal{N}$.
 \State Update time allocations of other CUs according to their proposals:
 \begin{equation}\label{a2}
 \hat{\alpha}^t_{m'n}=
 \begin{cases}
   \alpha^t_{m'}, & \mbox{if } b^t_{m'}=n \text{ and } \alpha^t_{m'}\neq \alpha_e  \\
   \hat{\alpha}^{t-1}_{m'n}, & \mbox{otherwise}.
 \end{cases}
 \end{equation}
 \State Update estimated utility with respect to joint D2D pair selection $\mathbf{b}=(n,\mathbf{b}_{-m})$, $\forall n\in\mathcal{N},\forall \mathbf{b}_{-m}\in\Pi_{m'\in\mathcal{M}\setminus\{m\}}\mathcal{B}_{m'}$:
 \begin{equation}\label{a3}
   \hat{\pi}_m(\mathbf{b})=
   \begin{cases}
     (1-\alpha^t_{mn})\hat{R}^C_{mn}(t)-{R}^C_m - \theta, & \mbox{if } Ch_{n}(\hat{\mathbf{b}})=m \\
     -\theta, & \mbox{otherwise},
   \end{cases}
 \end{equation}
 where $\hat{b}_{m}=(n,\alpha^t_{mn})$ and $\hat{b}_{m'}=(b_m,\hat{\alpha}^t_{m'b_{m'}})$ for $m'\neq m$.
 \State \textbf{End for}
 \end{algorithmic}
 }
 \end{algorithm}
 
\begin{theorem}
With $\varepsilon(t)=\varepsilon_0t^{-1/ML}$, the sequence $\{\alpha^t_{mn}\}$ converges to the true value $\alpha^*_{mn}$  with probability 1. Moreover, the algorithm converges to an SM  in probability. Specifically, $\lim_{t\rightarrow\infty}Pr\{\text{$\mu_{\mathbf{b}^t}$ is an SM}\}=1$, where $\mathbf{b}^t=(b^t_1,b^t_2,\cdots,b^t_M)$.
\end{theorem}

\begin{IEEEproof}
Let $P^t_{mn}$ denote the probability of CU $m$ cooperating with D2D pair $n$ at period $t$. Thus,
\begin{equation*}
  \sum_{t=1}^{\infty}P^t_{mn}\geq\sum_{t=1}^{\infty}\zeta\varepsilon(t)(1-\varepsilon(t))^{M-1}\geq\sum_{t=1}^{\infty}\frac{\zeta\varepsilon_0(1-\varepsilon_0)^{M-1}}{t^{1/ML}}=\infty.
\end{equation*}
So CU $m$ will cooperate with D2D pair $n$ infinitely often with probability 1. Based on \cite{chapman2013convergent}, $\left\{\hat{R}^C_{mn}(t)\right\}$ converges to ${R}^C_{mn}$ with probability 1. Since $\alpha^t_{mn}$ is a continuous function of $\hat{R}^C_{mn}(t)$, we conclude that $\left\{\alpha^t_{mn}\right\}$ converges to $\alpha^*_{mn}$ with probability 1.

\par On the one hand, if we replace the estimated utility $\hat{\pi}_m$ with  the true utility ${\pi}_m$ in EBRI-Q, the D2D pair selection process is exactly the stochastic BRI (SBRI) in \cite{chapman2013convergent}. Since $\mathcal{G}$ is a WABRG, using lemma 5.17 in \cite{chapman2013convergent}, we have that  $\lim_{t\rightarrow\infty}Pr\{\mathbf{b}^t \text{ is PNE}\}=1$ in this case. On the other hand, due to the step 3-a in EBRI-Q, the event that CU $m^\prime(m^\prime\neq m)$ announces its time allocation ${\alpha}^t_{m^\prime n}$ will happen infinitely often with probability 1.  So $\hat{\alpha}^t_{m^\prime n}$ converges to  ${\alpha}^*_{m^\prime n}$ with probability 1. Moreover, considering the convergence of $\hat{R}_{mn}(t)$ and $\alpha^t_{mn}$, the estimated utility will be sufficiently close to the true utility after an almost surely finite time. Thus, EBRI-Q will select D2D pairs with exactly the same probabilities as SBRI. Hence, based on Theorem 1, the convergence of $\mu_{\mathbf{b}^t}$ is verified.
\end{IEEEproof}

\begin{remark}
On the one hand, larger memory length $L$ improves the robustness to the exploration behavior of other D2D pairs, which may speed up the convergence rate of the algorithm. On the other hand, Theorem 3 implies that  the exploration probability $\varepsilon(t)$ decays more slowly with larger $L$, which leads to slower convergence rate.
\end{remark}

\subsection{Implementation Issues}
At the beginning of each frame, CUs will send their proposals to the BS. Then, the BS broadcasts a proposal list containing all the CUs' proposals at a dedicated channel. Meanwhile, all the D2D pairs will listen to this channel. After receiving CUs' proposals, each D2D pair will accept one of them. Then, each matched D2D pair will send a feedback to the BS using the channel occupied by its matched CU. Based on these feedback, the BS obtains the final matching and informs the result to the CUs. Thus, each CU knows its partner and can begin its data transmission. 

\par
Except the above hand-shaking procedure, no extra overhead is needed in the proposed algorithm.Thus, each iteration has low signaling overhead. Note that (\ref{a1})-(\ref{a3}) can be calculated in constant time. 
Moreover, the estimated utility is only needed in BRI. Thus, according to BRI, the algorithm only needs to estimate $\sum^{t-1}_{\tau=t-L}\pi_m(b,\mathbf{b}^{\tau}_{-m}),\forall b\in\mathcal{N}$. Therefore, the computational complexity of each iteration is $\mathcal{O}(LN)$.

\section{Simulation Results}
 Simulation results are presented to evaluate the performance of the proposed algorithm. The channel gain is $\eta D^{-K}$, where $D$ is the distance between receiver and transmitter, $K=4$ is the path loss exponent and $\eta$ is fast fading with exponential distribution. The cell radius is 400 m. CUs are randomly distributed in an area of at least 300 m away from the BS.  The distance between the DT and the BS is uniformly distributed between 150 and 250 m. The length of D2D link is uniformly distributed between 10 and 60 m. Besides, we set $n_0=-100$ dBm, $p_c=p_d=20$ mW, $\alpha_L=0.1$, $\alpha_U=0.5$, $\zeta=\varepsilon_0=0.1$ and $\xi=0.2$, and the length of memory in BRI is set to  4.

\par
\begin{figure}[!t]
\centering
\includegraphics[width=3.3in]{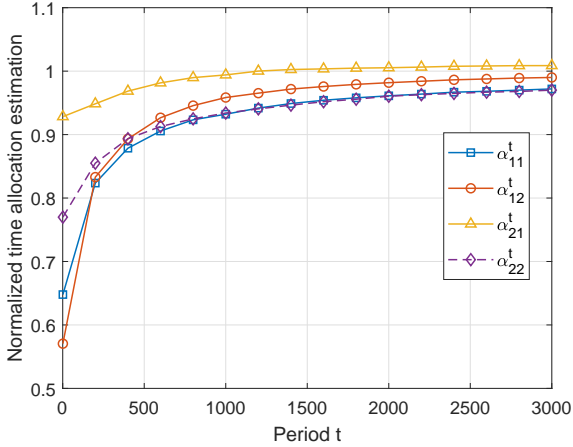}
\captionsetup{font={small}} 
\caption{Convergence of time allocation estimation with $M=2$, $N=2$.}
\label{convergence of time allocation}
\end{figure}
\begin{figure}[!t]
\centering
\subfloat[The behaviors of CU 1 over periods.]{\includegraphics[width=3.3 in]{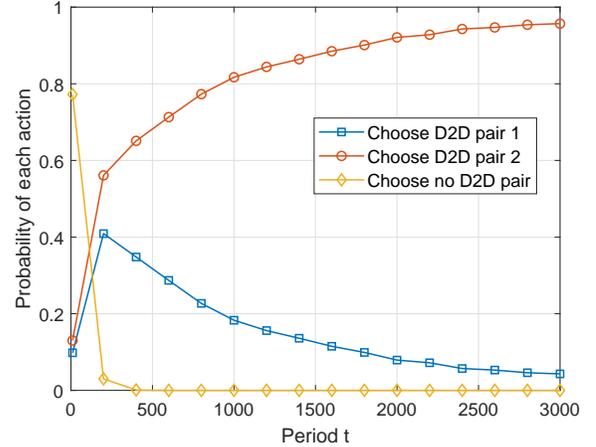}
\label{timeAllocation}}
\hfill
\subfloat[The behaviors of CU 2 over periods.]{\includegraphics[width=3.3 in]{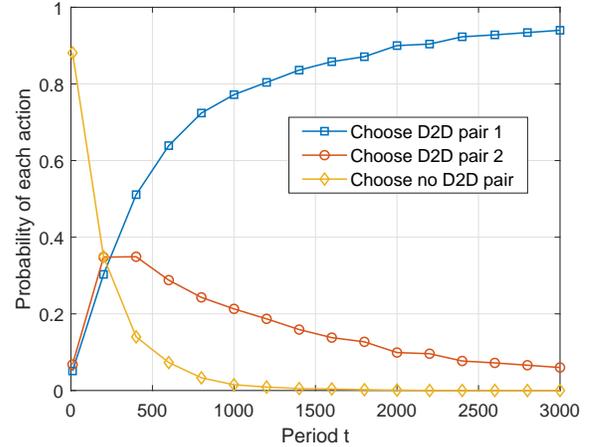}
\label{empiricalDist}}
\captionsetup{font={small}} 
\caption{The convergence of CUs' behaviors with $M=2$, $N=2$.}
\label{convergence of action}
\end{figure}

At first, we investigate the the convergence behavior of the proposed learning algorithm. For illustration purposes, we consider a small network with 2 CUs and 2 D2D pairs. There is only one SM, where CU 1 is matched with D2D pair 2 and CU 2 is matched with D2D pair 1. The results are given in Fig. \ref{convergence of time allocation} and Fig. \ref{convergence of action}. The results are averaged over 1000 simulations with the same topology.  Fig. \ref{convergence of time allocation} presents the convergence of the time allocation estimation, where the estimation $\alpha^t_{mn}$ is normalized by the true value $\alpha^*_{mn}$. It is observed that the sequence $\left\{\alpha^t_{mn}\right\}$ converges to $\alpha^*_{mn}$ asymptotically, which is consistent with Theorem 3. The convergence of CUs' behaviors is given in Fig. \ref{convergence of action}. It can be found that CU 1 and CU 2 could acquire their correct partners. This result implies that PNE or SM will be achieved eventually.

\begin{figure}[!t]
\centering
\includegraphics[width=3.3in]{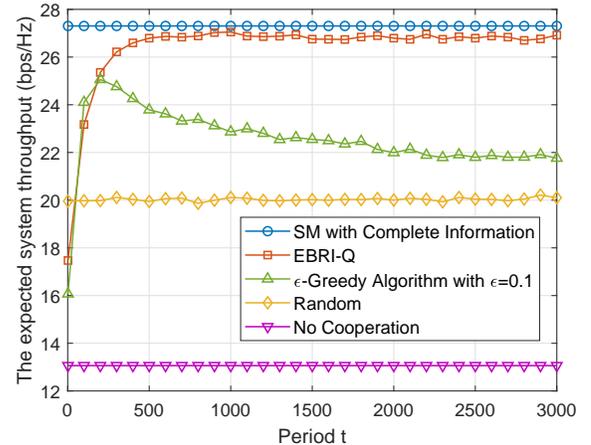}
\captionsetup{font={small}} 
\caption{Performance of the proposed algorithm compared to other algorithms with $M = 4, N=5$.}
\label{performCompare}
\end{figure}

Next, we compare the proposed algorithm with other  distributed algorithms in a larger network with 4 CUs and 5 D2D pairs. Fig. \ref{performCompare} shows the achieved system throughput over time for different algorithms. The results are averaged over 1000 simulations  with different topologies. In the classical exploration-exploitation $\epsilon$-greedy algorithm, at each period, every CU selects the best D2D pairs so far with probability $1-\epsilon$, and some random D2D pair with probability $\epsilon$. Besides, the time allocation estimations are updated similarly to our algorithm.  We take $\epsilon=0.1$ in the simulation. In the random algorithm, each CU selects D2D pair randomly and proposes $\alpha_L$ as time allocation to guarantee its performance. We present the non-cooperative scheme as well, where every CU takes \mbox{action $b_0$}. It can be observed that our algorithm yields significant gain over other learning algorithms. Besides, the performance loss due to incomplete information is small. It is also worth mentioning that the cooperative scheme achieves much better performance than non-cooperative scheme, which verifies the efficiency of the cooperative scheme.

\section{Conclusion}
This paper considers a cooperative D2D communication system with incomplete information. We model the pairing problem between multiple CUs and multiple D2D pairs as a one-to-one matching game and propose a novel learning algorithm, which converges to a stable matching. The simulation results verify our analysis and show that the proposed algorithm outperforms the classical $\epsilon$-greedy algorithm. In the future work, the location information will be considered to divide CUs and D2D pairs into small groups to speed up the learning process. Moreover, the learning algorithm with faster convergence rate will also be investigated.

\bibliographystyle{IEEEtran}
\bibliography{IEEEabrv,reference}

%

\ifCLASSOPTIONcaptionsoff
  \newpage
\fi

\end{document}